\documentclass[a4paper]{article}%
\usepackage{amsmath}
\usepackage{amsfonts}
\usepackage{amssymb}
\usepackage[lmargin=3cm,rmargin=3cm,tmargin=3cm,bmargin=3cm]{geometry}
\usepackage{graphicx}%

\begin{document}
\title{ Tau sleptons and Tau sneutrino Decays in MSSM
under the
 Cosmological Bounds }
\author{Levent Selbuz$^*$ and Z. Zekeriya Aydin$^\dag$ \\ \textit{Ankara University, Faculty of Engineering}\\ \textit{Department of Engineering Physics, 06100 Tandogan, Ankara, TURKEY}\\ \textit{$^*$selbuz@eng.ankara.edu.tr}\\ \textit{$^\dag$Z.Zekeriya.Aydin@eng.ankara.edu.tr} }

\maketitle
\begin{abstract}
We present the numerical investigation of the fermionic two-body
decays of tau sleptons $\tilde \tau_{1,2}$  and $\tau$ sneutrino in
the Minimal Supersymmetric Standard Model with complex parameters.
In the analysis we particularly take into account the cosmological
bounds imposed by WMAP data. We plot the CP-phase dependences for
each fermionic two-body channel of $\tilde \tau_{1,2}$ and $\tau$
sneutrino and speculate about the branching ratios and total
(two-body) decay widths. We find that the phase dependences of the
decay widths of the third family sleptons are quite significant
which can provide viable probes of additional CP sources. We also
draw attention to the polarization of the final-state tau in the $
\tilde \tau_{1,2}$ decays.
\end{abstract}

\section{Introduction}
The experimental HEP frontier is soon reaching TeV energies and most
of the physicists expect that just there theoretically proposed
Higgs bosons and superpartners are waiting to be discovered. There
are many reasons to be so optimistic. First of all, in spite of its
remarkable successes, the Standard Model has to be extended into a
more complete theory which should solve the hierarchy problem and
stabilize the Higgs boson mass against radiative corrections. The
most attractive extension to realize these objectives is
supersymmetry (SUSY) \cite{haber}. Its minimal version (MSSM)
requires a non-standard Higgs sector \cite{higgssector} which
introduces additional sources of
CP-violation\cite{Dugan:1984qf,Masiero:2002xj} beyond the
$\delta_{CKM}$ phase \cite{Cabibbo:yz}. The plethora of CP-phases
also influences the decays and mixings of B mesons (as well as D and
K mesons). The present experiments at BABAR, Tevatron and KEK and
the one to start at the LHC will be able to measure various decay
channels to determine if there are supersymmetric sources of CP
violation. In particular, CP-asymmetry and decay rate of $B
\rightarrow X_s \gamma$ form a good testing ground for low-energy
supersymmetry with CP violation \cite{bmeson}. The above-mentioned
additional CP-phases explain the cosmological baryon asymmetry of
the universe and the lightest SUSY particle could be an excellent
candidate for cold dark matter in the universe
\cite{Goldberg:1983nd,Ellis:1983ew}.

In the case of exact supersymmetry, all scalar particles would have
to have same masses with their associated SM partners. Since none of
the superpartners has been discovered, supersymmetry must be broken.
But in order to preserve the hierarchy problem solved the
supersymmetry must be broken softly. This leads to a reasonable mass
splittings between known particles and their superpartners, i.e. to
the superpartners masses around 1 TeV.

The precision experiments by Wilkinson Microwave Anisotropy Probe
(WMAP) \cite{Wilkinson} have put the following constraint on the
relic density of cold dark matter \footnote{In our calculation, we
have used WMAP-allowed bands in the plane $ M_1-\varphi$ which are
based on 1st year data. Now the WMAP 3rd year data is also available
\cite{Spergel:2006hy}, but the new WMAP + SDSS combined value for
relic density of dark matter does not change the numerical results
in Ref. \cite{Belanger}, namely the WMAP-allowed bands. See "Note
added" section of Ref. \cite{Belanger}.}
\begin{equation}\label{mkl0}
0.0945 < \Omega_{CDM}h^2 < 0.1287
\end{equation}

Recently, in the light of this cosmological constraint an extensive
analysis of the neutralino relic density in the presence of SUSY-CP
phases has been given by B\'{e}langer \textit{et al.}
\cite{Belanger}.

Analyses of the decays of third generation scalar quarks
\cite{Bartl} and scalar leptons \cite{Bartlslep} with complex SUSY
parameters have been performed by Bartl et al. In this study we
present the numerical investigation of the fermionic two-body decays
of third family sleptons in MSSM with complex SUSY parameters taking
into account the cosmological bound imposed by WMAP data. Actually,
we had performed some studies in this direction for squarks
\cite{Selbuz:2006fj,Aydin:2007aq} incorporating all the existing
bounds on the SUSY parameter space by utilizing the study by
Belanger et al. \cite{Belanger} before. These investigations showed
us that the effects of $M_1$ and its phase $\varphi_{U(1)}$ on the
decay widths of squarks are quite significant. Now we consider third
generation sleptons. Namely, we study the effect of $M_1$ and its
phase $ \varphi_{U(1)}$ on the decay widths of $\tilde \tau_{1,2}$
and $\tilde \nu_\tau$.

In the numerical calculations, although the SUSY parameters $\mu$,
$M_1$, $M_2$, and $A_f$ are in general complex, we assume that
$\mu$, $M_2$ and $A_f$ are real, but $M_1$ and its phase $
\varphi_{U(1)}$ take values on the WMAP- allowed bands given in Ref.
\cite{Belanger}. These bands also satisfy the EDM bounds
 \cite{edms}. The experimental upper limits on the EDMs of electron,
neutron and the $^{299}Hg$ and the $^{205}Tl$ atoms  may impose
constraints on the size of the SUSY CP-phases
\cite{Ellis:1982tk,Barger:2001nu}. However, these constraints are
highly model dependent.  This means that it is possible to suppress
the EDMs without requiring the various SUSY CP-phases be small. For
example, in the MSSM assuming strong cancellations between different
contributions \cite{Ibrahim:1997nc}, the phase of $\mu$ is
restricted to $|\varphi_\mu|< \pi/10$, but there is no such
restriction on the phases of $M_1$ and $A_f$. In addition, we
evaluate the parameter $M_2$ via the relation $
M_2=(3/5)|M_1|(\tan\theta_W)^{-2}$ which can be derived by assuming
gaugino mass unification purely in the electroweak sector of MSSM.
It is very important to insert the WMAP-allowed band in the plane $
M_1-\varphi$ into the numerical calculations instead of taking one
fixed $M_1$ value for all $\varphi$-phases, because, for example, on
the allowed band for $\mu=200$ GeV, $M_1$ starts from 140 GeV for
$\varphi=0$ and increasing monotonically it becomes 165 GeV for
$\varphi=\pi$. In Ref.\cite{Belanger} two WMAP-allowed band plots
are given, one for $\mu=200$ GeV and the other for $\mu=350$ GeV.
For both plots the other parameters are fixed to be $\tan \beta=10$,
$ m_{H^+}=1$ TeV, $A_f=1.2 $ TeV and
$\varphi_\mu$=$\varphi_{A_f}$=0. We here choose the masses for
$\tilde \tau_{1,2}$ sleptons as $m_{\tilde \tau_2}$=1000 GeV and
$m_{\tilde \tau_1}$=750 GeV. These $m_{\tilde \tau_{1,2}}$ values
lead to a sneutrino mass $m_{\tilde \nu_\tau}$=745 GeV for
$M_{\tilde L} < M_{\tilde E}$.

\section{Tau Sleptons and Tau Sneutrino Masses, Mixing and Decay Widths}
\subsection{Masses and mixing in slepton sector}
The superpartners of the SM fermions with left and right helicity
are the left and right sfermions. In the case of tau slepton (stau)
the left and right states are in general mixed. Therefore, the
sfermion mass terms of the Lagrangian are described in the basis
($\tilde \tau_{L}$,$\tilde \tau_{R}$) as
\cite{Ellis:1983ed,Gunion:1984yn}
\begin{equation}\label{mkl1}
 {\cal L}_M^{\tilde \tau }= -({\tilde \tau}_L^{\dag} {\tilde \tau}_R^{\dag})\left(
 \begin{array}{cc}
 M_{L L}^{2}& M_{L R}^{2}\\[1.ex]
 M_{R L}^{2} & M_{R R}^{2}
 \end{array}
 \right)
 \left(
\begin{array}{c}
\tilde \tau_L\\ [1.ex] \tilde \tau_R
\end{array}
\right)
\end{equation}
with
\begin{eqnarray}\label{mkl2}
M_{L L}^{2}&=&M_{\tilde
L}^{2}+(I_{3L}^{\tau}-e_\tau\sin^2\theta_W)\cos(2
\beta)m_{z}^{2}+m_{\tau}^{2},\\
M_{R R}^{2}&=&M_{\tilde E}^{2}+e_\tau\sin^2\theta_W\cos(2
\beta)m_{z}^{2}+m_{\tau}^{2},\\\label{mkl3} M_{R L}^{2}&=&(M_{L
R}^{2})^{*}=m_\tau(A_\tau-\mu^{*}(\tan\beta)^{-2I_{3L}^{\tau}}),\label{mkl4}
\end{eqnarray}
where $m_\tau$, $e_\tau$, $I_{3L}^{\tau}$ and $\theta_W$ are the
mass, electric charge, weak isospin of the $\tau$-lepton and the
weak mixing angle, respectively. $\tan\beta=v_2/v_1$ with $v_i$
being the vacuum expectation values of the Higgs fields $H_i^{0}$, $
i=1,2$. The soft SUSY-breaking parameters $M_{\tilde L}$, $M_{\tilde
E}$ and $A_\tau$  involved in Eqs. (3-5) can be evaluated for our
numerical calculations using the following relations:
\begin{eqnarray}\label{mkl5}
M_{\tilde L}^{2}&=&\frac{1}{2}{\left(m_{\tilde \tau_1}^{2}+m_{\tilde
\tau_2}^{2} \pm\sqrt{(m_{\tilde \tau_2}^{2}-m_{\tilde
\tau_1}^{2})^2-4m_\tau^{2}
|A_\tau-\mu^{*}\cot\beta|^2}\right)}\nonumber \\
&&+(\frac{1}{2}-\sin^2\theta_W)\cos(2\beta)m_{z}^{2}-m_{\tau}^{2},\\
\label{mkl6} M_{\tilde E}^{2}&=&\frac{1}{2}{\left(m_{\tilde
\tau_1}^{2}+m_{\tilde \tau_2}^{2} \mp\sqrt{(m_{\tilde
\tau_2}^{2}-m_{\tilde \tau_1}^{2})^2-4m_\tau^{2}
|A_\tau-\mu^{*}\cot\beta|^2}\right)}\nonumber \\
&&+\sin^2\theta_W\cos(2\beta)m_{z}^{2}-m_{\tau}^{2}
\end{eqnarray}

The $\tilde{\tau}$ mass eigenstates $\tilde \tau_1$ and $\tilde
\tau_2$ can be obtained from the weak states $\tilde \tau_L$ and
$\tilde \tau_R$ via the $\tilde \tau$-mixing matrix
\begin{equation}\label{mkl10}
 {\cal R}^{\tilde \tau }=\left(
 \begin{array}{cc}
 e^{i\varphi_{\tilde \tau}}\cos\theta_{\tilde \tau}& \sin\theta_{\tilde \tau}\\[1.ex]
 -\sin\theta_{\tilde \tau} & e^{-i\varphi_{\tilde \tau}}\cos\theta_{\tilde \tau}
 \end{array}
 \right)
\end{equation}
where
 \begin{equation}\label{mkl9}
 \varphi_{\tilde \tau}=\arg[M_{R
 L}^{2}]=\arg[A_\tau-\mu^{*}(\tan\beta)^{-2I_{3L}^{\tau}}]
 \end{equation}
 and
\begin{equation}\label{mkl11}
 \cos\theta_{\tilde \tau}=\frac{-|M_{L R}^{2}|}
 {\sqrt{|M_{L R}^{2}|^2+
 (m_{\tilde \tau_1}^{2}-M_{L L}^{2})^2}}, \qquad
\sin\theta_{\tilde \tau}=\frac{M_{L L}^{2}-m_{\tilde \tau_1}^{2}}
 {\sqrt{|M_{L R}^{2}|^2+
 (m_{\tilde \tau_1}^{2}-M_{L L}^{2})^2}}
\end{equation}
One can easily get the following stau mass eigenvalues by
diagonalizing the mass matrix in Eq. (2):
\begin{equation}\label{mkl12}
 m_{\tilde \tau_{1,2}}^{2}=\frac{1}{2}
{\left(M_{L L}^{2}+M_{R R}^{2} \mp\sqrt{(M_{L L}^{2}-M_{R
R}^{2})^2+4|M_{L R }^{2}|^2} \right)} ,\qquad m_{\tilde \tau_1}<
m_{\tilde \tau_2}
\end{equation}

The $\tilde \nu_\tau $ appears only in the left state. Its mass is
given by
\begin{equation}\label{mkl12a}
 m_{\tilde \nu_\tau}^{2}=M_{\tilde L}^{2}+\frac{1}{2}\cos(2\beta)m_z^2
\end{equation}

Note that in this work we neglect CP-violation effects related to
flavor change. Besides that the scalar mass matrices and trilinear
scalar coupling parameters are assumed to be flavor diagonal.

\subsection{Fermionic decay widths of  $\tilde \tau_i $ and  $\tilde \nu_\tau $ }

The lepton-slepton-chargino and lepton-slepton-neutralino
Lagrangians have been first given in Ref. 1. Here we use them in
notations of Ref. 13:
\begin{eqnarray}\label{mkl13}
{\cal L}_{l' \tilde l \tilde \chi^{\pm}}=g\bar{u}(\ell_{i j}^{\tilde
d}P_R + k_{i j}^{\tilde d}P_L){\tilde \chi}_j^{+}{\tilde
d}_i+g\bar{d}(\ell_{i j}^{\tilde u}P_R + k_{i j}^{\tilde
u}P_L){\tilde \chi}_j^{+c}{\tilde u}_i+h.c.
\end{eqnarray}
and
\begin{eqnarray}\label{mkl18}
{\cal L}_{l \tilde l \tilde \chi^{0}} =g\bar{l}(a_{i j}^{\tilde
l}P_R + b_{i j}^{\tilde l}P_L){\tilde \chi}_j^{0}{\tilde l}_i+h.c.
\end{eqnarray}
where u ($\tilde u$) stands for (s)neutrinos and d ($\tilde d$)
stands for charged (s)leptons. We also borrow the formulas for the
partial decay widths of $\tilde l_i$ ($\tilde l_i$ = $\tilde \tau_i$
and $\tilde \nu_\tau$) into  lepton-neutralino (or chargino) from
Ref. 13. The partial decay width for the decay $\tilde
\tau_i\rightarrow \tilde \chi_j^{0}+\tau(\lambda_\tau)$ is expressed
as

\begin{eqnarray}\label{mkl22}
\Gamma(\tilde \tau_i\rightarrow \tilde
\chi_j^{0}+\tau(\lambda_\tau))&=&\frac{g^2\kappa^{1/2}( m_{\tilde
\tau_{i}}^2,m_{\tilde \chi_j^{0}}^2,m_{\tau}^{2})}{16\pi m_{\tilde
\tau_{i}}^3}|\mathcal{M}_{\lambda_\tau}|^2
\end{eqnarray}
with
\begin{eqnarray}\label{mkl23}
|\mathcal{M}_{\lambda_\tau}|^2&=&\frac{1}{4}\{H_{s}^2[|b_{ij}^{\tilde
\tau}|^2
+|a_{ij}^{\tilde \tau}|^2+2Re(b_{ij}^{\tilde \tau * }a_{ij}^{\tilde \tau})]\nonumber\\
&& + H_{p}^2[|b_{ij}^{\tilde \tau}|^2 +|a_{ij}^{\tilde
\tau}|^2-2Re(b_{ij}^{\tilde \tau * }a_{ij}^{\tilde \tau})]\nonumber\\
&& + 2(-1)^{\lambda_\tau+(1/2)}H_{p}H_{s}(|a_{ij}^{\tilde
\tau}|^2-|b_{ij}^{\tilde \tau}|^2)\}
\end{eqnarray}
where $\lambda_{\tau}=\pm\frac{1}{2}$ is the helicity of the
outgoing $\tau$, $\kappa(x,y,z)=x^{2}+y^{2}+z^{2}-2(xy+xz+yz)$,
$H_{s}=[m_{\tilde \tau_{i}}^2-(m_{\tilde
\chi_j^{0}}+m_{\tau})^2]^{1/2}$ and $H_{p}=[m_{\tilde
\tau_{i}}^2-(m_{\tilde \chi_j^{0}}-m_{\tau})^2]^{1/2}$.

The explicit forms of the couplings, $a_{i j}^{\tilde \tau}$,  $b_{i
j}^{\tilde \tau}$ and $\ell_{i j}^{\tilde \tau}$, are

\begin{equation}\label{mkl151}
a_{i j}^{\tilde \tau}={\cal R}_{in}^{{\tilde \tau}^{*}}{\cal
A}_{jn}^{\tau}, \qquad   b_{i j}^{\tilde \tau}={\cal
R}_{in}^{{\tilde \tau}^{*}}{\cal B}_{jn}^{\tau},\qquad\ell_{i
j}^{\tilde \tau}={\cal R}_{in}^{{\tilde \tau}^{*}} {\cal
O}_{jn}^{\tau} \qquad (n=L,R)
\end{equation}
where
\begin{equation}
{\cal A}_j^{\tau}=
\begin{pmatrix}
\ f_{Lj}^\tau \\ \ h_{Rj}^\tau\\
\end{pmatrix}, \qquad
{\cal B}_j^{\tau}=
\begin{pmatrix}
\ h_{Lj}^\tau \\ \ f_{Rj}^\tau\\
\end{pmatrix}, \qquad
{\cal O}_j^{\tau}=
\begin{pmatrix}
\ -U_{j1} \\ \ Y_{\tau}U_{j2}\\
\end{pmatrix},
\end{equation}
and
\begin{eqnarray}\label{mkl20}
f_{Lj}^{\tau}&=&-\frac{1}{\sqrt{2}}(N_{j2}+\tan\theta_WN_{j1})\nonumber \\
f_{Rj}^{\tau}&=&\sqrt{2}\tan\theta_WN_{j1}^{*}\nonumber \\
h_{Lj}^{\tau}&=&(h_{Rj}^{\tau})^{*}=Y_{\tau}N_{j3}^{*}.
\end{eqnarray}

The partial decay width of  $\tilde \tau_i$ into the chargino,
$\tilde \tau_i\rightarrow \tilde \chi_j^{-}+\nu{_\tau}$, is obtained
by the replacements $a_{ij}^{\tilde \tau}\rightarrow \ell_{i
j}^{\tilde \tau}$, $b_{ij}^{\tilde \tau}\rightarrow0$, $m_{\tilde
\chi_j^{0}}\rightarrow m_{\tilde \chi_j^{-}}$,
$m_{\tau}\rightarrow0$ and $\lambda_{\tau}\rightarrow-\frac{1}{2}$
in Eq. (15) and Eq. (16) with the couplings $\ell_{i j}^{\tilde
\tau}$ also given in Eq. (17) and Eq. (18).

The width for the  $\tau$-sneutrino decay
$\tilde\nu_{\tau}\rightarrow{\tilde\chi_j^{0}}\nu_{\tau}$ is
obtained by the replacements $a_{ij}^{\tilde \tau}\rightarrow
a_{j}^{\tilde \nu}$, $b_{ij}^{\tilde \tau}\rightarrow 0$, $m_{\tilde
\tau_{i}}\rightarrow m_{\tilde \nu_{\tau}}$, $m_{\tau}\rightarrow0$
and $\lambda_{\tau}\rightarrow-\frac{1}{2}$ in Eq. (15) and Eq.
(16), and that for the decay
$\tilde\nu_{\tau}\rightarrow{\tilde\chi_j^{+}}\tau(\lambda_\tau)$ by
the replacements $a_{ij}^{\tilde \tau}\rightarrow \ell_{j}^{\tilde
\nu}$, $b_{ij}^{\tilde \tau}\rightarrow k_{j}^{\tilde \nu}$,
$m_{\tilde \tau_{i}}\rightarrow m_{\tilde \nu_{\tau}}$ and
$m_{\tilde \chi_j^{0}}\rightarrow m_{\tilde \chi_j^{+}}$. The
coupling are now

\begin{equation}\label{mkl15d}
a_{j}^{\tilde \nu}=\frac{1}{\sqrt{2}}(\tan\theta_WN_{j1}-N_{j2}),
\qquad   k_{j}^{\tilde \nu}=Y_{\tau}U_{j2}^{*}, \qquad
\ell_{j}^{\tilde \nu}=-V_{j1}.
\end{equation}

Here, N is the $4\times4$ neutralino mixing matrix, U and V are
$2\times2$ chargino mixing matrices and
$Y_{\tau}=m_{\tau}/(\sqrt{2}m_{W}\cos\beta)$ is the $\tau$ Yukawa
coupling.

In this work we contend with tree-level amplitudes as we aim at
determining the phase-sensitivities of the decay rates, mainly.

\section{Tau-Slepton and Tau-Sneutrino Decays}

Here we present the dependences of the $\tilde \tau_{1,2}$ and
$\tilde \nu_{\tau}$ two-body decay widths on the $ \varphi_{U(1)}$
for $ \mu =200 $ GeV  and for $ \mu =350 $ GeV. We also choose the
values for the masses ($m_{\tilde \tau_1}$, $m_{\tilde \tau_2}$,
$m_{\tilde \chi_1^\pm}$, $m_{\tilde \chi_2^\pm}$, $m_{\tilde
\chi_1^0}$) = (750 GeV, 1000 GeV,
 180 GeV, 336 GeV, 150 GeV) for  $ \mu =200 $ GeV and ($m_{\tilde \tau_1}$, $m_{\tilde \tau_2}$, $m_{\tilde \chi_1^\pm}$,
$m_{\tilde \chi_2^\pm}$, $m_{\tilde \chi_1^0}$) = (750 GeV, 1000
GeV, 340 GeV, 680 GeV, 290 GeV) for $ \mu =350 $ GeV. The mass
values of these $\tau$-sleptons lead to a sneutrino mass $m_{\tilde
\nu_\tau}$=745 GeV. Note that although the neutralino and chargino
masses vary with $\varphi_{U(1)}$, these variations are not large.
Therefore, as a final state particle (i.e., on mass-shell), we have
chosen fixed (average) mass values for charginos and neutralinos.
For both sets of values by calculating the  $M_{\tilde L}$ and
$M_{\tilde E}$ values corresponding to $m_{\tilde \tau_1}$ and
$m_{\tilde \tau_2}$, we plot the decay widths  for $M_{\tilde L}
\geq M_{\tilde E}$  and $M_{\tilde L} < M_{\tilde E}$, separately.
We plot the $\varphi_{U(1)}$-dependences of the $\tilde \nu_{\tau}$
partial decay widths only for $M_{\tilde L} < M_{\tilde E}$. In the
case of $M_{\tilde L} > M_{\tilde E}$, the phase dependences do not
change, but decay widths take larger values. In our figures, we
display the slepton decay widths for the both helicity states of the
outgoing $\tau$ ($\tau_{L}$ and $\tau_{R}$).

\underline{{\textit{$M_{\tilde L} > M_{\tilde E}$ for $ \mu =200 $
GeV:}}}

In Figure 1(a) we show the partial decay widths of the channels $
\tilde \tau_1\rightarrow \tilde \chi_1^- \nu_{\tau}
 $, $ \tilde \tau_1\rightarrow \tilde \chi_2^- \nu_{\tau}
 $, $ \tilde \tau_1\rightarrow \tilde \chi_1^0 \tau_{L,R}
 $, $ \tilde \tau_2\rightarrow \tilde \chi_1^- \nu_{\tau}
 $, $ \tilde \tau_2\rightarrow \tilde \chi_2^- \nu_{\tau}
 $ and $ \tilde \tau_2\rightarrow \tilde \chi_1^0 \tau_{L,R}
 $  as a function of $ \varphi_{U(1)}$ for $ \mu =200$
 GeV. In these plots some dependences on the $ \varphi_{U(1)}$ phase are shown. In order
to see these dependences more pronouncedly, we now plot two channels
separately; namely $\tilde \tau_1\rightarrow \tilde \chi_1^0
\tau_{R}$ (the variations in the cross section are not large) and $
\tilde \tau_2\rightarrow \tilde \chi_1^0 \tau_{L}$ (the variations
are really large) in Figures 4(a)-(b). Here we consider the case
$M_{\tilde L} > M_{\tilde E}$, where $ \tilde \tau_1$ is mainly $
\tilde \tau_R$-like and $ \tilde \tau_2$ is mainly $ \tilde
\tau_L$-like (${\cal R}_{11}^{\tilde \tau }$=${\cal R}_{22}^{\tilde
\tau }$$\approx$ 0). In this case, the decay processes whose initial
and final state helicities are the same, $ \tilde \tau_2\rightarrow
\tilde \chi_2^- \nu_{\tau} $, $ \tilde \tau_2\rightarrow \tilde
\chi_1^- \nu_{\tau}$, $\tilde \tau_2\rightarrow \tilde \chi_1^0
\tau_{L}
 $ and $\tilde \tau_1\rightarrow \tilde \chi_1^0 \tau_{R}
 $, have large widths, whereas those with opposite helicities, $\tilde \tau_2\rightarrow \tilde \chi_1^0
 \tau_{R}$, $\tilde \tau_1\rightarrow \tilde \chi_1^- \nu_{\tau}$, $\tilde \tau_1\rightarrow \tilde \chi_2^-
 \nu_{\tau}$ and $ \tilde \tau_1\rightarrow \tilde \chi_1^0
 \tau_{L}$, have small ones. The reason for these large and small widths
can be traced to the couplings $a_{ij}^{\tilde \tau}$,
$b_{ij}^{\tilde \tau}$, $\ell_{i j}^{\tilde \tau}$ and
$a_{j}^{\tilde \nu}$, $k_{j}^{\tilde \nu}$, $\ell_{j}^{\tilde \nu}$.
Because of $H_{s}$$\approx$$H_{p}$ (since $m_{\tilde
 \tau_{1,2}}$$\gg$$m_{\tau}$) we can express the decay widths of
$\tilde \tau_i\rightarrow \tilde \chi_j^0 \tau(\lambda_\tau)$ as $
\Gamma(\tilde \tau_i\rightarrow \tilde \chi_j^0
\tau(\lambda_\tau=1/2))$$\propto$$|b_{ij}^{\tilde \tau}|^2$ and $
\Gamma(\tilde \tau_i\rightarrow \tilde \chi_j^0
\tau(\lambda_\tau=-1/2))$$\propto$$|a_{ij}^{\tilde \tau}|^2$. For
example, $ \Gamma(\tilde \tau_1\rightarrow \tilde \chi_1^0
\tau_{L})$ ($ \Gamma(\tilde \tau_2\rightarrow \tilde \chi_1^0
\tau_{R})$) is suppressed because it is proportional to the term,
$|a_{11}^{\tilde \tau}|$ $(|b_{21}^{\tilde \tau}|)$, which includes
small Yukawa coupling ($Y_{\tau}$). On the other hand, $
\Gamma(\tilde \tau_1\rightarrow \tilde \chi_1^0 \tau_{R})$ is
proportional to the square of $b_{11}^{\tilde \tau}$  which depends
on the combination ${\cal R}_{11}^{{\tilde \tau}^{*}}{\cal
B}_{11}^{\tau}+{\cal R}_{12}^{{\tilde \tau}^{*}}{\cal
B}_{12}^{\tau}$ contributing largely from its second term.
Similarly, since $H_{s}$=$H_{p}$, the decay widths of $\tilde
\tau_i\rightarrow \tilde\chi_j^{-}+\nu{_\tau}$ can be expressed as
$\Gamma(\tilde \tau_i\rightarrow
\tilde\chi_j^{-}+\nu{_\tau})$$\propto$$H_s^2$$|\ell_{ij}^{\tilde
\tau}|^2$. The decay widths of $ \tilde \tau_1\rightarrow \tilde
\chi_1^- \nu_{\tau}
 $, $ \tilde
\tau_1\rightarrow \tilde \chi_2^- \nu_{\tau}$ are also  suppressed
due to the very small Yukawa coupling ($\ell_{11}^{\tilde
\tau}$$\thickapprox$$Y_{\tau}$ ${{\cal
R}_{12}^{\tilde\tau}}^{*}U_{12}$, $\ell_{12}^{\tilde
\tau}$$\thickapprox$$Y_{\tau}$ ${{\cal
R}_{12}^{\tilde\tau}}^{*}U_{22}$).

Note that the decay width $\Gamma(\tilde \tau_1\rightarrow \tilde
\chi_1^0 \tau_{R})$ is 90-110 times larger than $\Gamma(\tilde
\tau_1\rightarrow \tilde \chi_1^0 \tau_{L})$ and $\Gamma(\tilde
\tau_2\rightarrow \tilde \chi_1^0 \tau_{L})$ is 10-30 times larger
than $\Gamma(\tilde \tau_2\rightarrow \tilde \chi_1^0 \tau_{R})$.
From Figure 1(a) one can see that the branching ratios for $ \tilde
\tau_2$ are roughly $B( \tilde \tau_2\rightarrow \tilde \chi_2^-
\nu_{\tau})$ :
 $B(\tilde \tau_2\rightarrow \tilde \chi_1^0 \tau_{L})$ :
$B(\tilde \tau_2\rightarrow \tilde \chi_1^- \nu_{\tau})$ : $B(\tilde
\tau_2\rightarrow \tilde \chi_1^0 \tau_{R})$  $\approx$  6 : 1 : 0.5
: 0.03.

\begin{figure}
[t]
\begin{center}
\includegraphics[height=5cm,width=8cm]{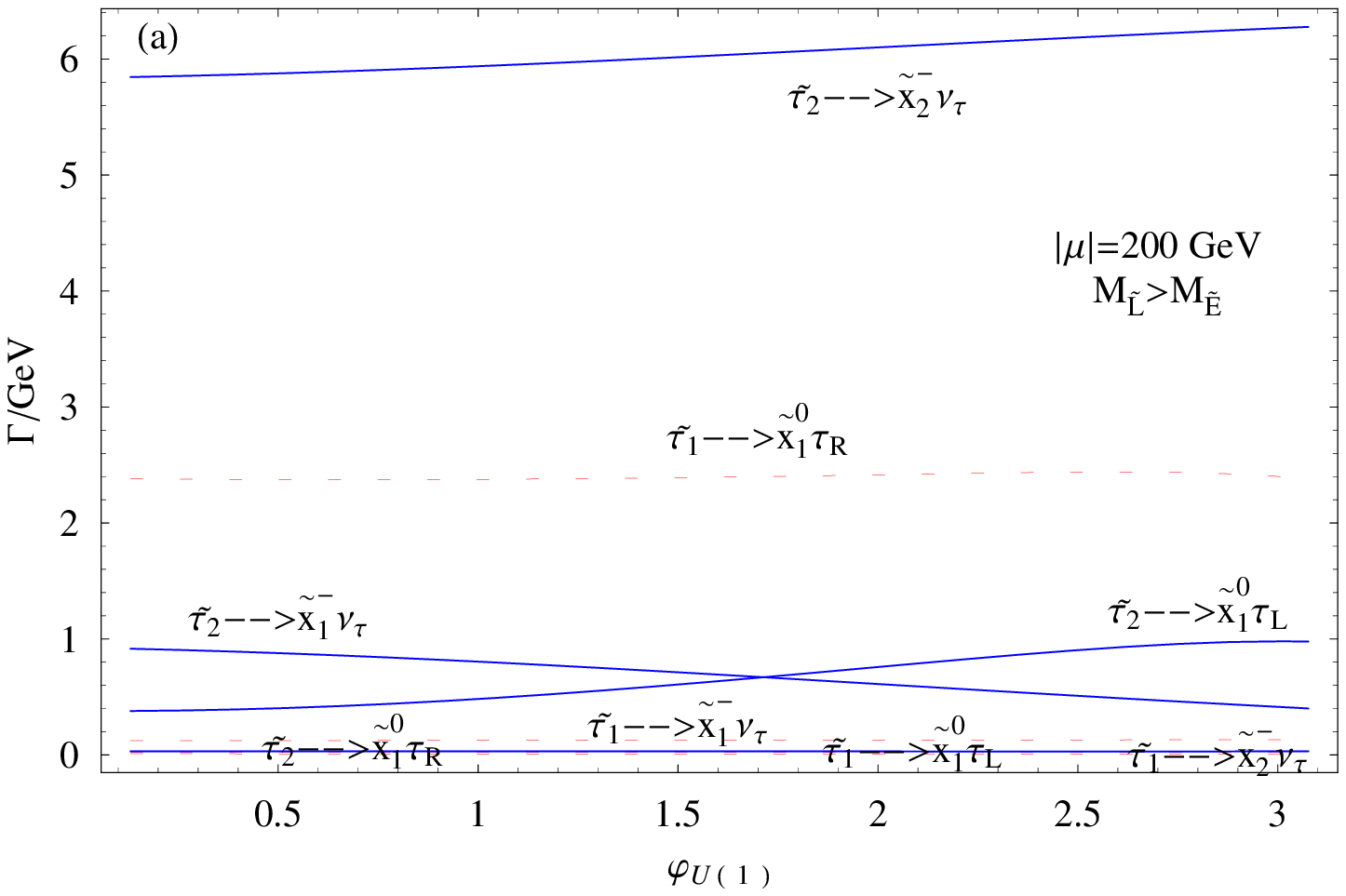}
\hspace{1cm}%
  \includegraphics[height=5cm,width=8cm]{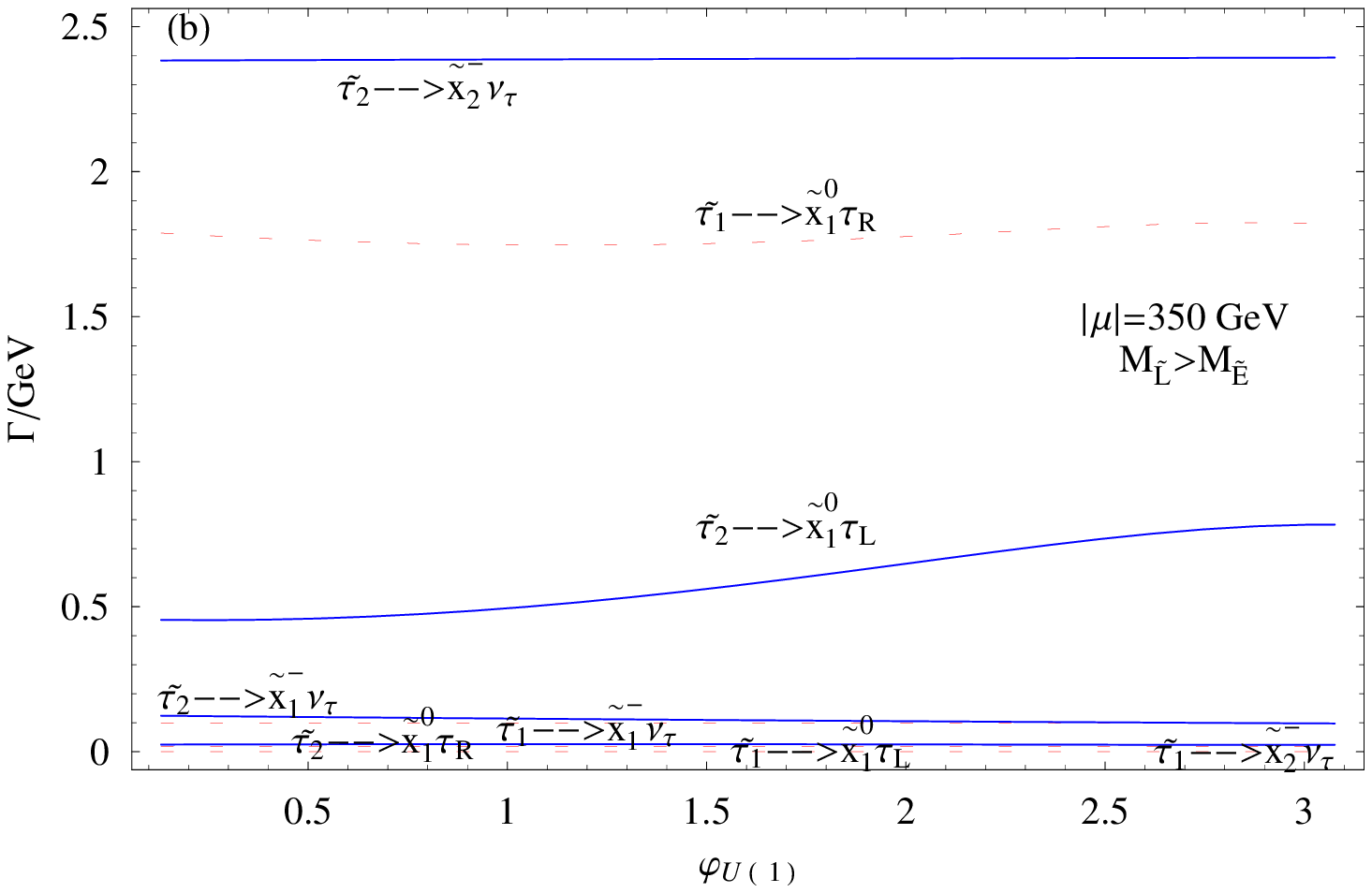}
  \label{dosfiguras13523}
  \hspace{1cm}%
\caption{Partial decay widths $\Gamma$ of the  $\tilde \tau_{1,2}$
decays for $\tan\beta=10$, $A_\tau=1.2 $ TeV,
$\varphi_\mu$=$\varphi_{A_\tau}$=0, $m_{\tilde \tau_1}=750$ GeV,
$m_{\tilde \tau_2}=1000$ GeV and $M_{\tilde L} > M_{\tilde E}$; $\mu
=200 $ GeV in (a) and $\mu =350 $ GeV in (b).}
\label{fig:dosfiguras233}
\end{center}
\end{figure}

Although the $ \varphi_{U(1)}$ dependence of $\Gamma(\tilde
\tau_2\rightarrow \tilde \chi_1^0 \tau_{L}
 )$ ($\Gamma(\tilde \tau_1\rightarrow \tilde \chi_1^0 \tau_{R}
 )$) stems only from the parameters $|N_{11}|$ and $|N_{12}|$ ($|N_{11}|$), the phase dependence is quite pronounced.
Similarly, the $ \varphi_{U(1)}$ phase dependence of $ \tilde
\tau_2\rightarrow \tilde \chi_{1}^-(\tilde \chi_{2}^-) \nu_{\tau} $
stemmed only from the $ \varphi_{U(1)}$ dependence of $|U_{11}|$
($|U_{21}|$) parameter is also quite pronounced. The decay width
$\Gamma(\tilde \tau_1\rightarrow \tilde \chi_1^0 \tau_{R})$ takes
its maximum (minimum) value at
$\varphi_{U(1)}$$\approx$$\frac{5\pi}{6}$
($\varphi_{U(1)}$$\approx$$\frac{\pi}{4}$) (see Figure 4(a)). This
$\varphi_{U(1)}$ value also corresponds to maximum (minimum) value
of $|b_{11}^{\tilde \tau}|^2$. In a similar way, the width
$\Gamma(\tilde \tau_2\rightarrow \tilde \chi_1^0 \tau_{L}
 )$ and its parameter $|a_{21}^{\tilde \tau}|^2$ takes their
 maximum (minimum) value at $\varphi_{U(1)}$$\approx$$\pi$ ($\varphi_{U(1)}$=0) (see Figure 4(b)). Hence, we can say
that the phase $\varphi_{U(1)}$ dependence of $|a_{i j}^{\tilde
\tau}|^2$ and $|b_{i j}^{\tilde \tau}|^2$ ($|\ell_{i j}^{\tilde
\tau}|^2$) reflects the phase $\varphi_{U(1)}$ dependence of
channels $\tilde \tau_i\rightarrow \tilde \chi_j^0 \tau_{R,L}$
($\tilde \tau_i\rightarrow \tilde \chi_j^-\nu_{\tau}$) directly.

\begin{figure}
[t]
\begin{center}
  \includegraphics[height=5cm,width=8cm]{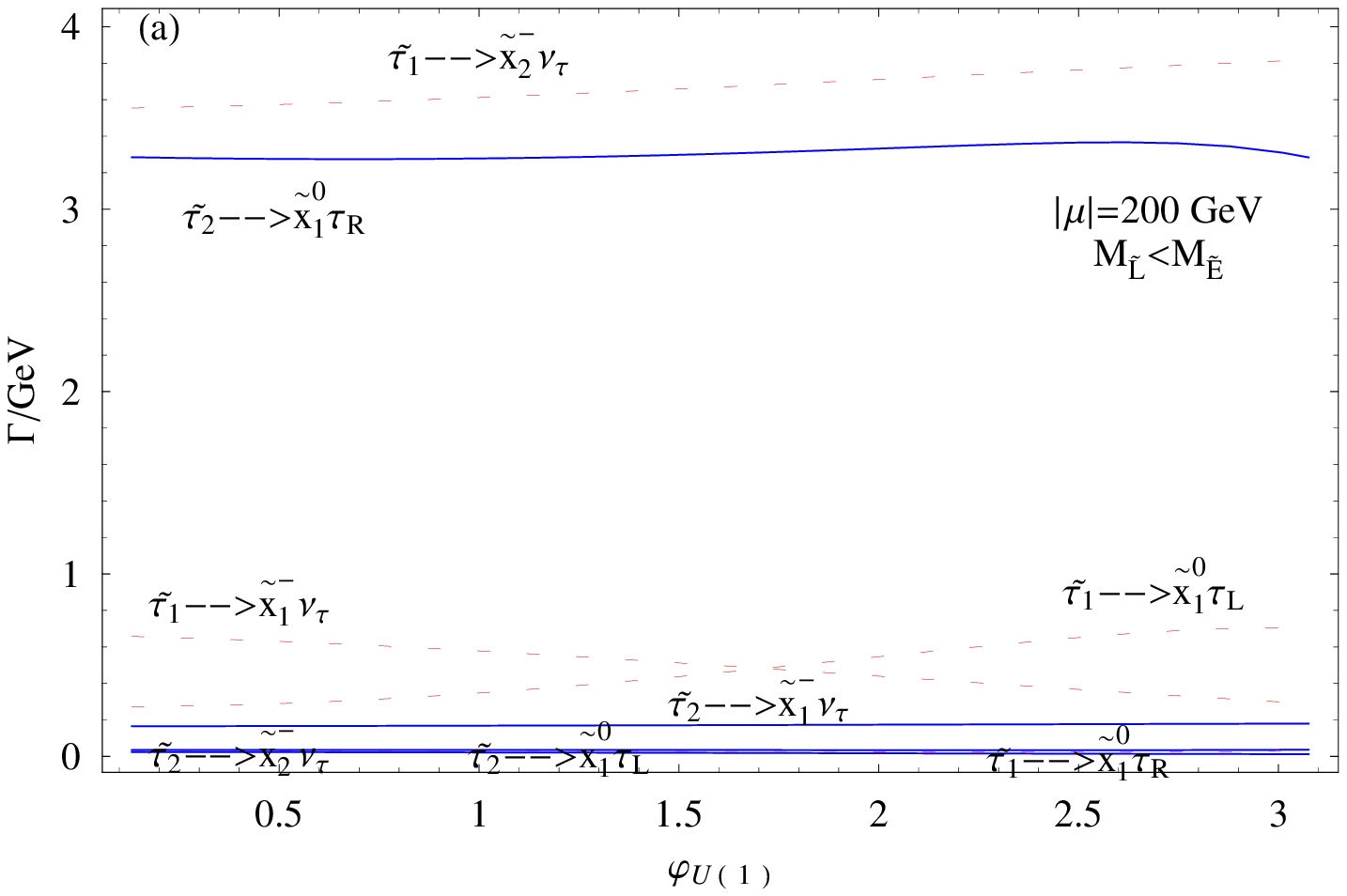}
\hspace{1cm}%
  \includegraphics[height=5cm,width=8cm]{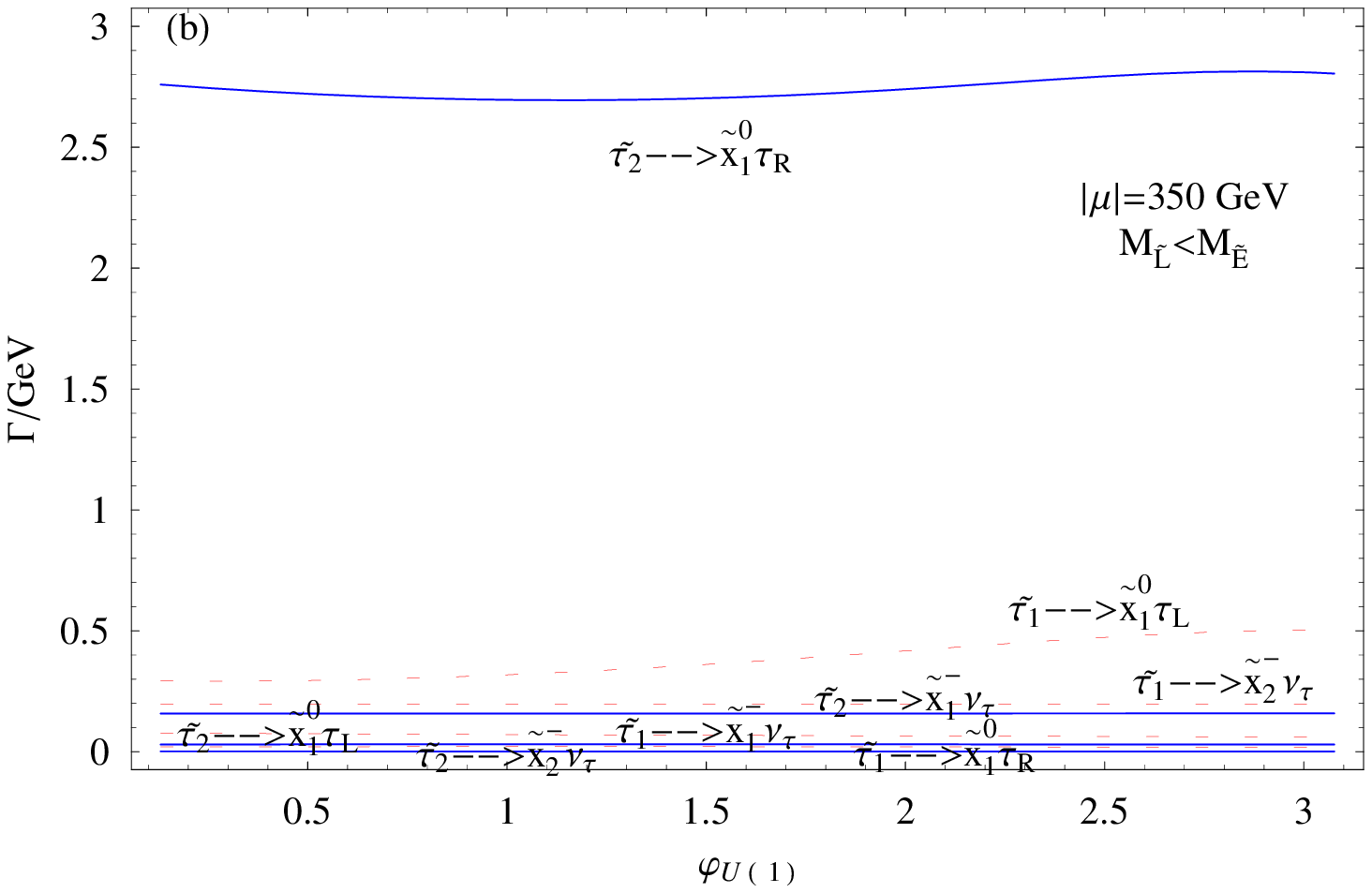}
  \label{dosfiguras1673}
  \hspace{1cm}%
\caption{Partial decay widths $\Gamma$ of the  $\tilde \tau_{1,2}$
decays for $\tan\beta=10$, $A_\tau=1.2 $ TeV,
$\varphi_\mu$=$\varphi_{A_\tau}$=0, $m_{\tilde \tau_1}=750$ GeV,
$m_{\tilde \tau_2}=1000$ GeV and $M_{\tilde L} < M_{\tilde E}$; $\mu
=200 $ GeV in (a) and $\mu =350 $ GeV in (b).}
  \label{dosfiguras893}
\end{center}
\end{figure}

\underline{\textit{$M_{\tilde L} > M_{\tilde E}$ for $ \mu =350 $
GeV :}}

We give the same partial decay widths in Figure 1(b) for $\mu =350$
GeV (See also Figures 4(c)-(d)). Here, too, $ \tilde \tau_1$ is
mainly $ \tilde \tau_R$-like and $ \tilde \tau_2$ is mainly $ \tilde
\tau_L$-like because we still keep the case $M_{\tilde L}> M_{\tilde
E}$. For  $ \mu =350 $ GeV the WMAP-allowed band \cite{Belanger}
takes place in larger $M_1$ values ($ \sim 305-325$ GeV) leading to
larger chargino and neutralino masses. This leads to smaller
$H_s^2$(since $H_s^2$ $\propto$ $[m_{\tilde \tau_{i}}^2-m_{\tilde
\chi_j}^2]$) and, as a result, smaller widths for $\tilde
\tau_{1,2}$ decays compared with those for $ \mu =200$ GeV.

As can be seen from Figure 4(d), the $\varphi_{U(1)}$ dependence of
the decay $\tilde \tau_2\rightarrow \tilde \chi_1^0 \tau_{L}$ is
prominent such that the value of decay width at
$\varphi_{U(1)}$=$\pi$ is about 2 times larger than that at
$\varphi_{U(1)}$=$0$. The decay widths $\Gamma(\tilde
\tau_2\rightarrow \tilde \chi_1^0 \tau_{R})$, $\Gamma(\tilde
\tau_1\rightarrow \tilde \chi_1^- \nu_{\tau})$, $\Gamma(\tilde
\tau_1\rightarrow \tilde \chi_2^- \nu_{\tau})$ and $\Gamma(\tilde
\tau_1\rightarrow \tilde \chi_1^0 \tau_{L})$  are suppressed because
of the same reasons mentioned above. The decay width of the process
$\tilde \tau_2\rightarrow \tilde \chi_2^- \nu_{\tau}$ is the largest
one among the $\tilde \tau_2$ channels and the branching ratios are
$B(\tilde \tau_2\rightarrow \tilde \chi_2^- \nu_{\tau})$ :
 $B(\tilde \tau_2\rightarrow \tilde \chi_1^0 \tau_{L})$ :
$B(\tilde \tau_2\rightarrow \tilde \chi_1^- \nu_{\tau})$ : $B(\tilde
\tau_2\rightarrow \tilde \chi_1^0 \tau_{R})$  $\approx$  2.4 : 0.8 :
0.1 : 0.02.

\underline{\textit{$M_{\tilde L} < M_{\tilde E}$ for $ \mu =200 $
GeV: }}

We give $\tilde \tau_{1,2}$ and $\tilde \nu_{\tau}$ decay widths as
a function of $\varphi_{U(1)}$ in Figure 2(a) and Figure 3(a)
respectively (for $ \mu =200$ GeV). In Figures 4(e)-(f) we plot two
of them separately whose CP-phase dependences are not clearly seen
in Figure 2(a). They, too, show the significant dependences on
CP-violation phase. In this subsection we consider the case
$M_{\tilde L} < M_{\tilde E}$, where $ \tilde \tau_1$ is mainly $
\tilde \tau_L$-like and $ \tilde \tau_2$ is mainly $ \tilde
\tau_R$-like (${\cal R}_{12}^{\tilde \tau }$=${\cal R}_{21}^{\tilde
\tau }$$\approx$0). The decay width of the process $\tilde
\tau_1\rightarrow \tilde \chi_2^- \nu_{\tau}$ is the largest one
among the $\tilde \tau_{1,2}$ channels; its decay width increases
from 3.55 GeV to 3.8 GeV monotonically as $\varphi_{U(1)}$ increases
from 0 to $\pi$. In this case ($M_{\tilde L} < M_{\tilde E}$), the
width $\Gamma(\tilde \tau_1\rightarrow \tilde \chi_2^- \nu_{\tau})$
is not suppressed because its $\ell_{12}^{\tilde \tau}$ term does
not include  Yukawa coupling ($\ell_{12}^{\tilde
\tau}$$\thickapprox$ ${{\cal R}_{11}^{\tilde\tau}}^{*}U_{21}$). The
phase dependence of $\tilde \tau_2\rightarrow \tilde \chi_1^0
\tau_{R}$ can be seen clearly in Figure 4(f); $\Gamma(\tilde
\tau_2\rightarrow \tilde \chi_1^0 \tau_{R})$ takes its minimum and
maximum values at $\varphi_{U(1)}$$\approx$$\frac{\pi}{4}$ and at
$\varphi_{U(1)}$$\approx$$\frac{5\pi}{6}$ respectively, because the
parameter $|b_{21}^{\tilde \tau}|^2$ reaches its minimum and maximum
at these $\varphi_{U(1)}$ values.

\begin{figure}
[t]
\begin{center}
  \includegraphics[height=5cm,width=8.06cm]{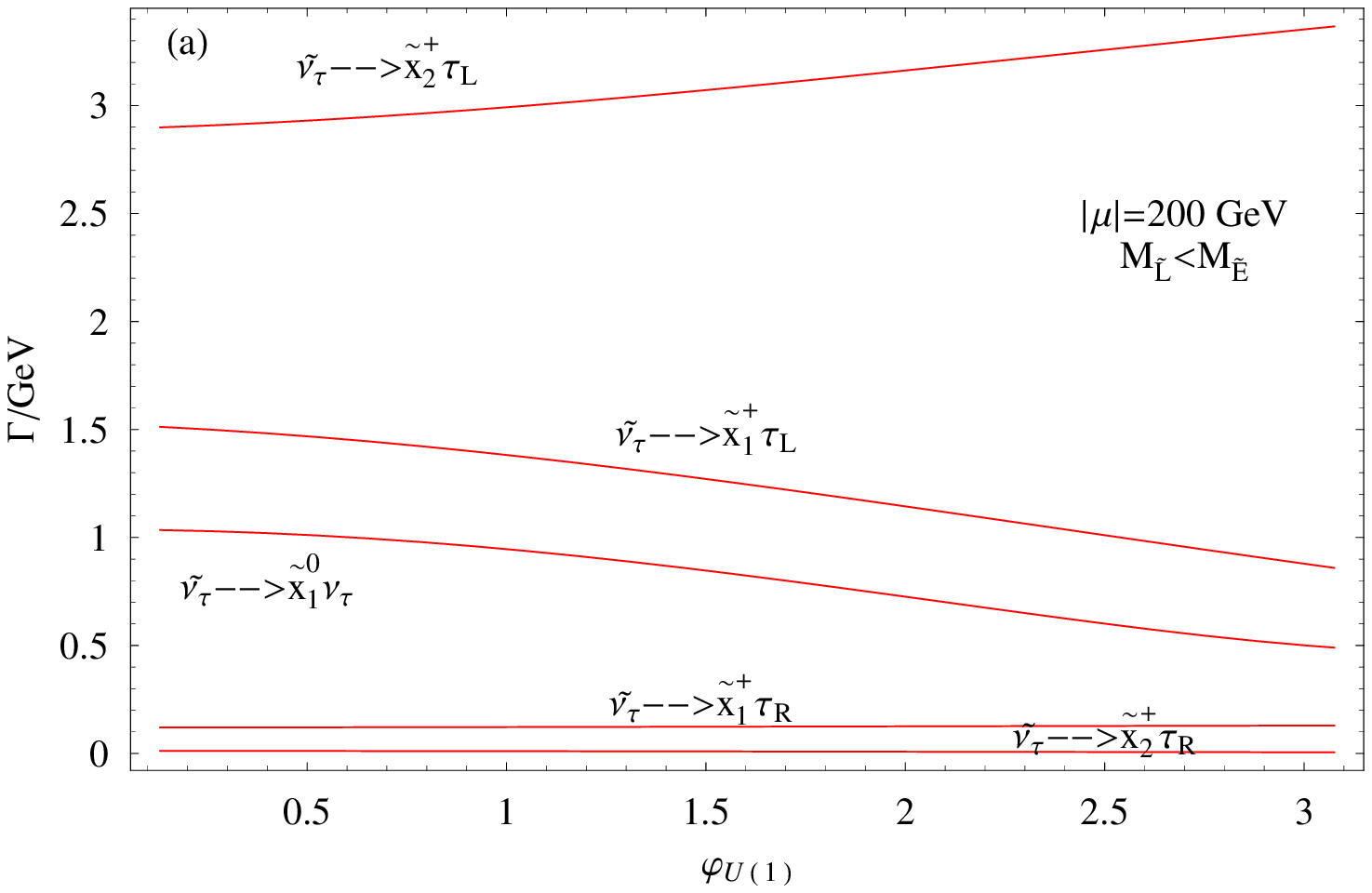}
\hspace{1cm}%
  \includegraphics[height=5cm,width=8.1cm]{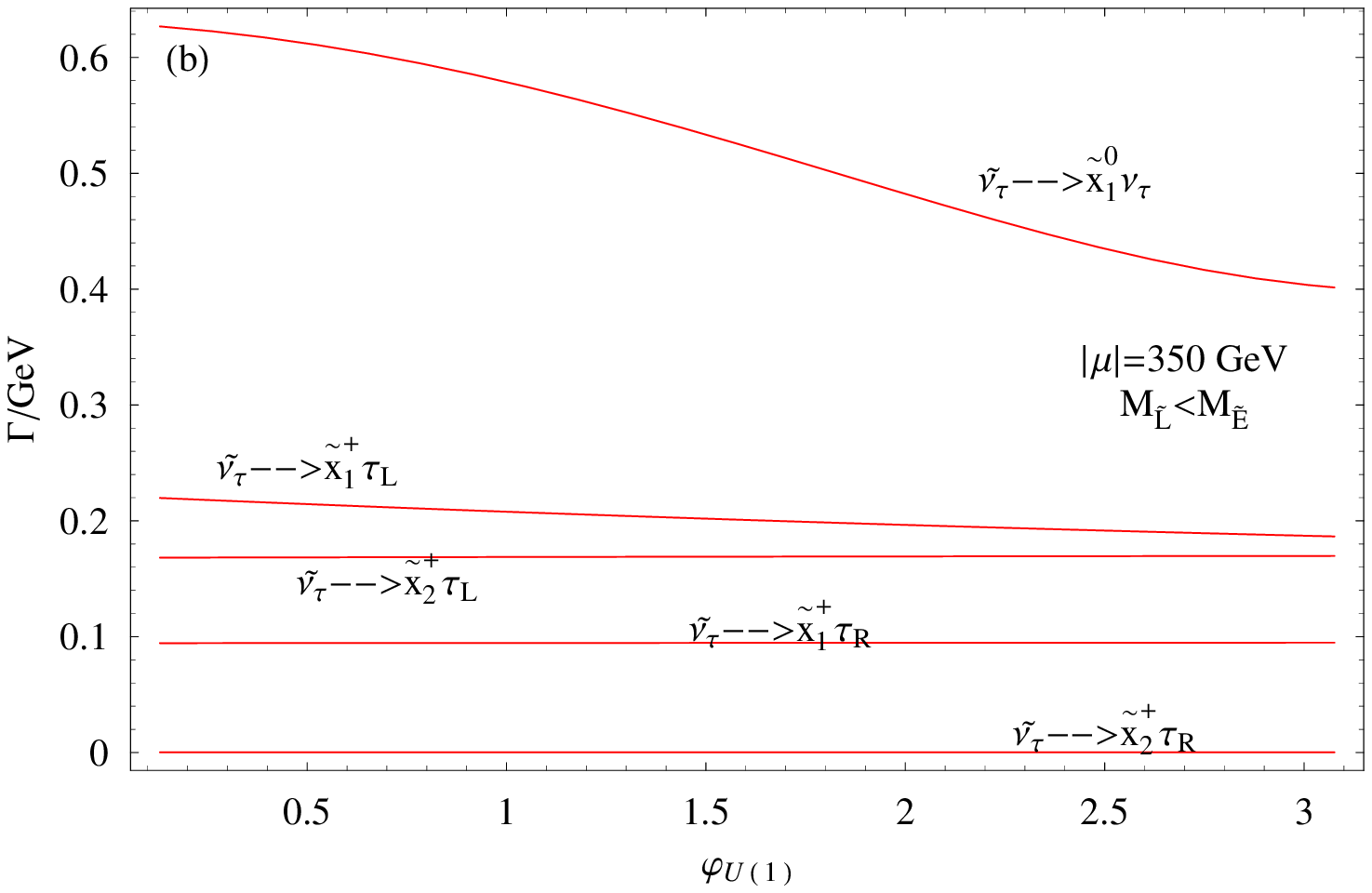}
  \label{dosfiguras16578}
  \hspace{1cm}%
\caption{Partial decay widths $\Gamma$ of the  $\tilde \nu_\tau$
decays for $\tan\beta=10$, $A_\tau=1.2 $ TeV,
$\varphi_\mu$=$\varphi_{A_\tau}$=0, $m_{\tilde \tau_1}=750$ GeV,
$m_{\tilde \tau_2}=1000$ GeV and $m_{\tilde \nu_\tau}=745$ GeV; $\mu
=200 $ GeV in (a) and $\mu =350 $ GeV in (b).}
  \label{dosfiguras85673}
\end{center}
\end{figure}

The branching ratios for $ \tilde \tau_1$ decays are roughly $B(
\tilde \tau_1\rightarrow \tilde \chi_2^- \nu_{\tau})$ :
 $B(\tilde \tau_1\rightarrow \tilde \chi_1^0 \tau_{L})$ :
$B(\tilde \tau_1\rightarrow \tilde \chi_1^- \nu_{\tau})$ : $B(\tilde
\tau_1\rightarrow \tilde \chi_1^0 \tau_{R})$  $\approx$  3.8 : 0.7 :
0.3 : 0.02.

In Figure 3(a) we give $\tilde \nu_{\tau}$ decay widths as a
function of $ \varphi_{U(1)}$ for $ \mu =200 $ GeV. The phase
dependence is more significant for the decay channels $\tilde
\nu_{\tau}\rightarrow \tilde \chi_2^+ \tau_{L}$, $\tilde
\nu_{\tau}\rightarrow \tilde \chi_1^+ \tau_{L}$ and $\tilde
\nu_{\tau}\rightarrow \tilde \chi_1^0\nu_{\tau}$. Analogously to the
neutralino decays of $\tilde \tau_{1,2}$; because of
$H_{s}$$\approx$$H_{p}$ (since $m_{\tilde
\nu_{\tau}}$$\gg$$m_{\tau}$) we can express the decay widths of
$\tilde \nu_{\tau}\rightarrow \tilde \chi_j^+ \tau(\lambda_\tau)$ as
$\Gamma(\tilde \nu_{\tau}\rightarrow \tilde \chi_j^0
\tau(\lambda_\tau=1/2))$$\propto$$|k_{j}^{\tilde \nu}|^2$ and
$\Gamma(\tilde \nu_{\tau}\rightarrow \tilde \chi_j^0
\tau(\lambda_\tau=-1/2))$$\propto$$|\ell_{j}^{\tilde \nu}|^2$. To be
more specific, $ \Gamma(\tilde \nu_{\tau}\rightarrow \tilde \chi_1^+
\tau_{R})$ ($ \Gamma(\tilde \nu_{\tau}\rightarrow \tilde \chi_2^+
\tau_{R})$) is suppressed because it is proportional to the term
$|k_{1}^{\tilde \nu}|$ $(|k_{2}^{\tilde \nu}|)$ which includes small
Yukawa coupling ($Y_{\tau}$). Since $H_{s}$=$H_{p}$ for neutrinos,
the decay width of $\tilde \nu_{\tau}\rightarrow \tilde
\chi_1^0+\nu_\tau$ can be expressed as $\Gamma(\tilde
\nu_{\tau}\rightarrow \tilde
\chi_1^0+\nu_\tau)$$\propto$$H_s^2$$|a_1^{\tilde\nu}|^2$. The $
\varphi_{U(1)}$ dependences of $\Gamma(\tilde \nu_{\tau}\rightarrow
\tilde \chi_j^+ \tau_{L})$ ($\Gamma(\tilde \nu_{\tau}\rightarrow
\tilde \chi_1^0\nu_\tau)$) stems from the $ \varphi_{U(1)}$
dependences of $|\ell_{j}^{\tilde \nu}|$ ($|a_1^{\tilde\nu}|$)
parameter and this parameter is quite phase-dependent.

The branching ratios for $\tilde \nu_{\tau}$ decays are roughly
$B(\tilde \nu_{\tau}\rightarrow \tilde \chi_2^+ \tau_{L})$ :
$B(\tilde \nu_{\tau}\rightarrow \tilde \chi_1^+ \tau_{L})$ :
$B(\tilde \nu_{\tau}\rightarrow \tilde \chi_1^0\nu_{\tau})$ :
$B(\tilde \nu_{\tau}\rightarrow \tilde \chi_1^+ \tau_{R})$ :
$B(\tilde \nu_{\tau}\rightarrow \tilde \chi_2^+ \tau_{R})$ $\approx$
3 : 1.3 : 0.1 : 0.01.

\begin{figure}
  \centering
\includegraphics[width=0.46\textwidth]{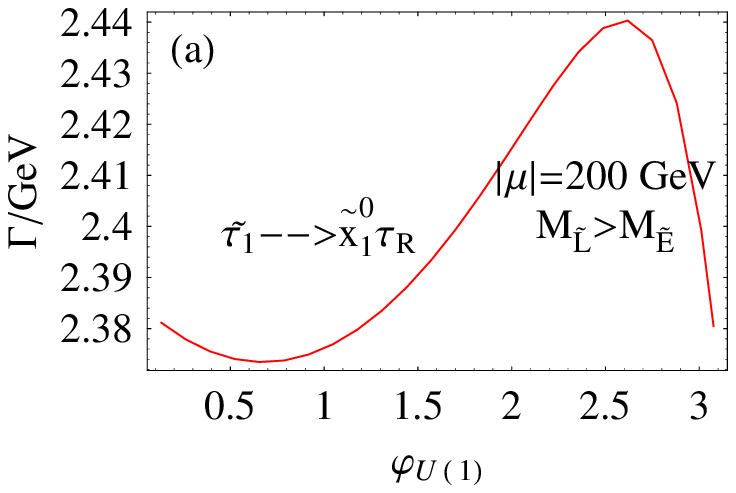}
\hspace{1cm}%
  \includegraphics[width=0.46\textwidth]{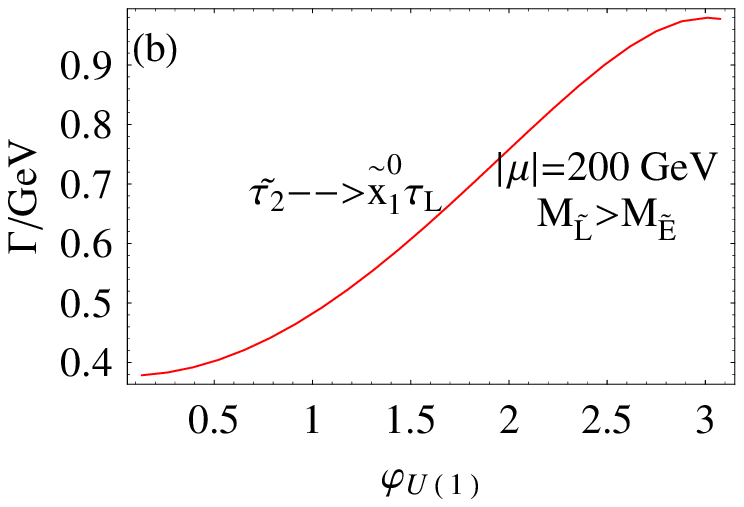}
  \label{dosfiguras1323}
  \centering
\includegraphics[width=0.46\textwidth]{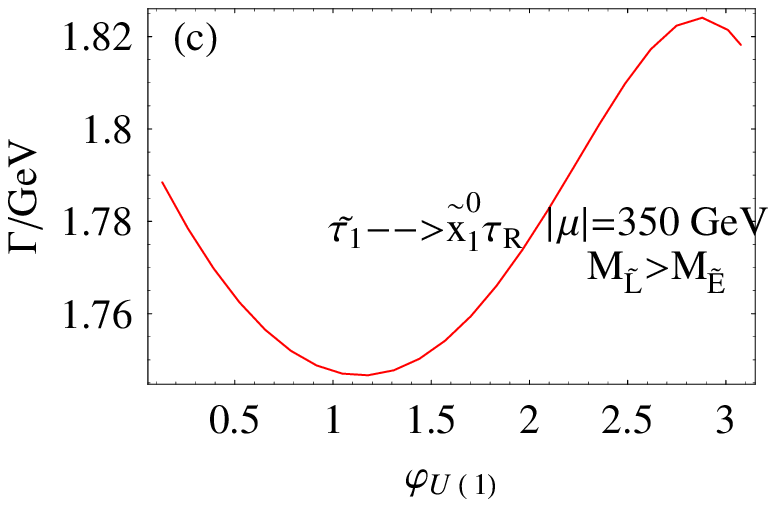}
 \hspace{1cm}%
  \includegraphics[width=0.46\textwidth]{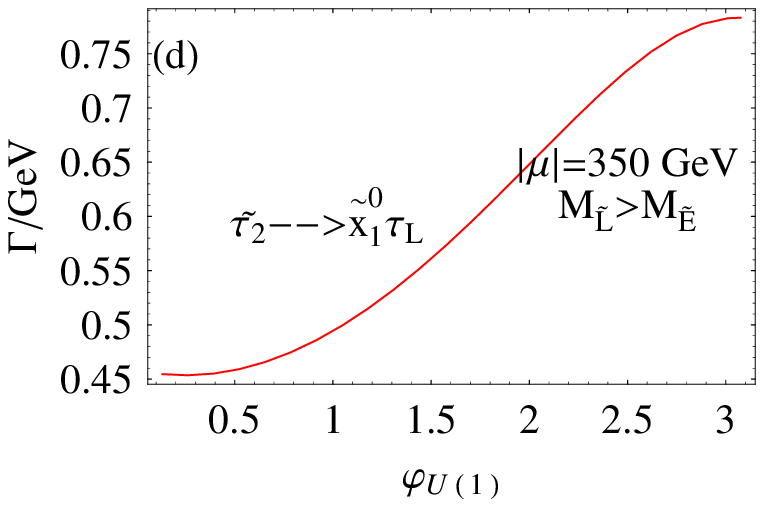}
  \label{dosfiguras134}
  \centering
\includegraphics[width=0.46\textwidth]{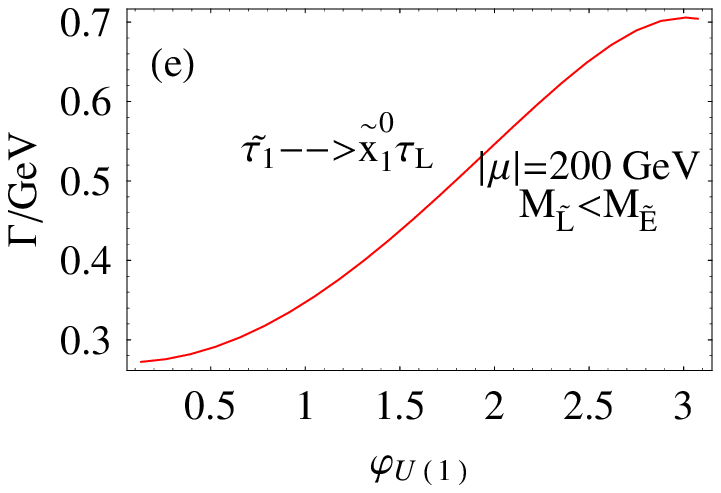}
\hspace{1cm}%
   \includegraphics[width=0.46\textwidth]{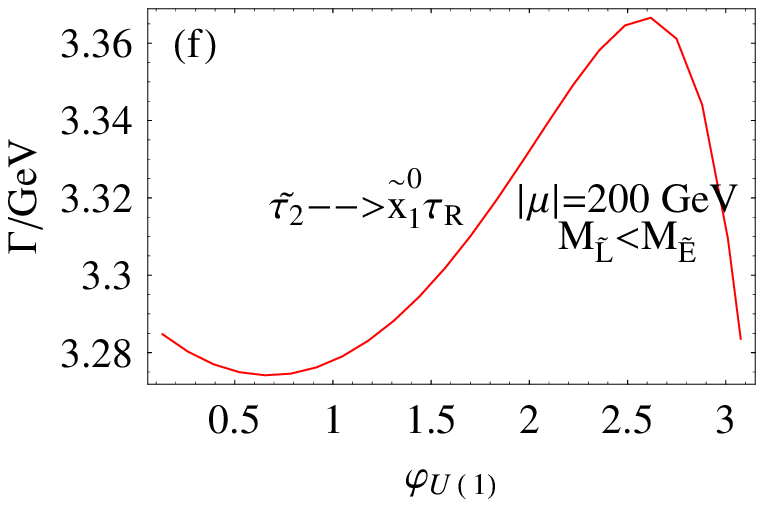}
\centering
\includegraphics[width=0.46\textwidth]{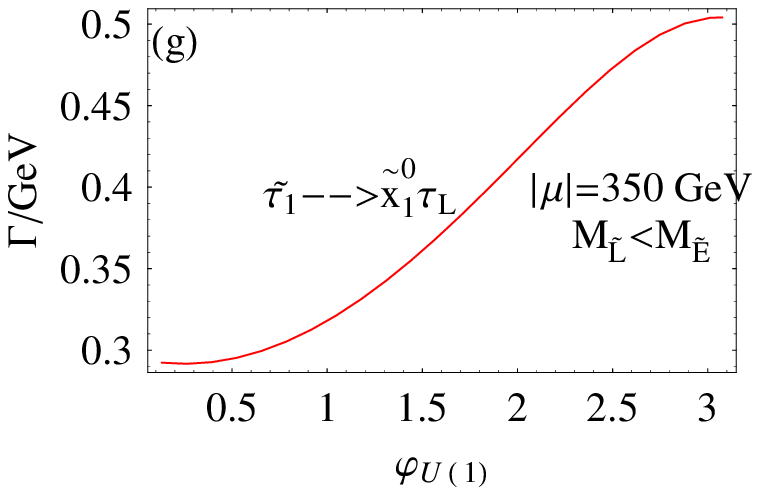}
\hspace{1cm}%
  \includegraphics[width=0.46\textwidth]{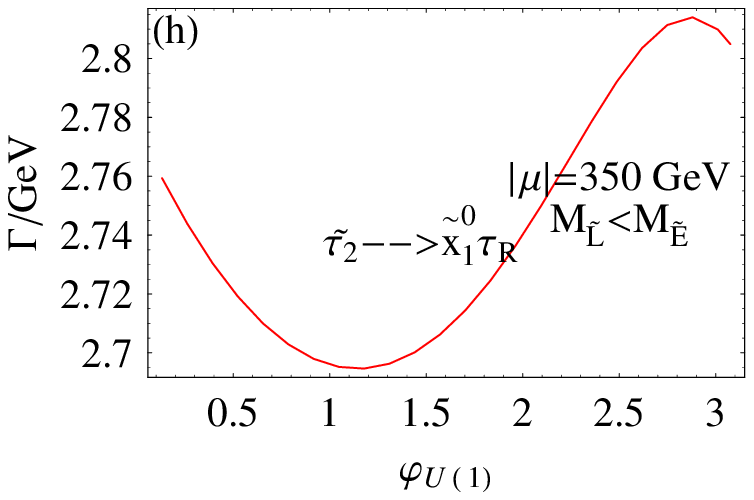}
  \label{dosfiguras1393}
\hspace{1cm}%
\caption{(a)-(h)  $ \varphi_{U(1)}$ dependences of certain $\tilde
\tau_{1,2}$ decays for $\mu =200$ GeV and for $\mu =350$ GeV.}
  \label{dosfiguras233}
\end{figure}

\underline{\textit{$M_{\tilde L} < M_{\tilde E}$ for $ \mu =350 $
GeV :}}

We present the dependences of the $\tilde \tau_{1,2}$ and $\tilde
\nu_{\tau}$ partial decay widths on $\varphi_{U(1)}$ in Figure 2(b)
and Figure 3(b) (for $ \mu =350 $ GeV). In this case
$H_s^2$($H_p^2$) takes smaller value because of the reason mentioned
in the previous subsection and this leads to smaller widths for
$\tilde \tau_{1,2}$ and $\tilde \nu_{\tau}$. In Figures 4(g)-(h) we
again plot two $\tilde \tau_{1,2}$ decay channels separately whose
phase dependences are not clearly seen in Figure 2(b). The
dependence of the phase $\varphi_{U(1)}$ in $\tilde \tau_{1,2}$
decays  are similar to those in the case $M_{\tilde L} < M_{\tilde
E}$ (for $ \mu =200$ GeV).

Note that $\Gamma(\tilde \tau_2\rightarrow \tilde \chi_1^0
\tau_{R})$$\approx$ 90 $\Gamma(\tilde \tau_2\rightarrow \tilde
\chi_1^0 \tau_{L})$ and $\Gamma(\tilde \tau_1\rightarrow \tilde
\chi_1^0 \tau_{L})$$\approx$ 30 $\Gamma(\tilde \tau_1\rightarrow
\tilde \chi_1^0 \tau_{R})$.

The width $\Gamma (\tilde \nu_{\tau}\rightarrow \tilde
\chi_1^0+\nu_\tau)$ decreases as the phase increases from 0 to
$\pi$, showing a significant dependence on the phase. The branching
ratios are roughly $B(\tilde \nu_{\tau}\rightarrow \tilde
\chi_1^0\nu_{\tau})$ : $B(\tilde \nu_{\tau}\rightarrow \tilde
\chi_1^+ \tau_{L})$ : $B(\tilde \nu_{\tau}\rightarrow \tilde
\chi_2^+ \tau_{L})$ : $B(\tilde \nu_{\tau}\rightarrow \tilde
\chi_1^+ \tau_{R})$ : $B(\tilde \nu_{\tau}\rightarrow \tilde
\chi_2^+ \tau_{R})$ $\approx$ 0.6 : 0.2 : 0.15 : 0.09 : 0.0002.

\section{Discussion and Summary  }
In this paper, we have presented the numerical investigation of the
fermionic two-body decays of third family sleptons in the minimal
supersymmetric standard model with complex parameters taking into
account the cosmological bounds imposed by WMAP data. For this
purpose, we have calculated numerically the decay widths of tau
sleptons $\tilde \tau_{1,2}$  and $\tau$ sneutrino, paying
particular attention to their dependence on the CP phase $
\varphi_{U(1)}$. We have found that some decay channels like $\tilde
\tau_2\rightarrow \tilde \chi_2^- \nu_{\tau}$, $\tilde
\tau_2\rightarrow \tilde \chi_1^- \nu_{\tau}$, $\tilde
\tau_2\rightarrow \tilde \chi_1^0 \tau_{L}$, $\tilde
\tau_2\rightarrow \tilde \chi_1^0 \tau_{R}$, $\tilde
\tau_1\rightarrow \tilde \chi_2^- \nu_{\tau}$, $\tilde
\tau_1\rightarrow \tilde \chi_1^- \nu_{\tau}$, $\tilde
\tau_1\rightarrow \tilde \chi_1^0 \tau_{L}$, $\tilde
\nu_{\tau}\rightarrow \tilde \chi_2^+ \tau_{L}$, $\tilde
\nu_{\tau}\rightarrow \tilde \chi_1^+ \tau_{L}$ and $\tilde
\nu_{\tau}\rightarrow \tilde \chi_1^0\nu_{\tau}$ show considerable
dependences on $ \varphi_{U(1)}$ phase. These decay modes will be
observable at a future $\emph{e}^+\emph{e}^-$ collider and LHC.
Therefore they provide viable probes of CP violation beyond the
simple CKM framework; moreover, they carry important information
about the mechanism that brakes Supersymmetry.

Besides that, $\tilde \tau$ decay is important since it is the sole
process where one can get information of the sfermion mixing and the
neutralino mixing from the polarization of the final-state fermion
\cite{Nojiri:1994it}. Note indeed that for $ \mu =200 $ GeV and
$M_{\tilde L} > M_{\tilde E}$ the decay width $\Gamma(\tilde
\tau_1\rightarrow \tilde \chi_1^0 \tau_{R})$ is 90-110 times larger
than $\Gamma(\tilde \tau_1\rightarrow \tilde \chi_1^0 \tau_{L})$ and
$\Gamma(\tilde \tau_2\rightarrow \tilde \chi_1^0 \tau_{L})$ is 10-30
times larger than $\Gamma(\tilde \tau_2\rightarrow \tilde \chi_1^0
\tau_{R})$ since $\tilde \tau_1$ ($\tilde \tau_2$) is mainly $\tilde
\tau_R$-like ( $\tilde \tau_L$-like). For $ \mu =200 $ GeV and
$M_{\tilde L} < M_{\tilde E}$ make only the interchange $\tilde
\tau_{1}\leftrightarrow\tilde \tau_{2}$ everywhere in the
above-mentioned preceding sentence. For $ \mu =350 $ GeV the pattern
expressed above remains more or less the same.

The phase dependence of the fermionic two-body decay widths of
$\tilde \tau_i$ and $\tilde\nu_i$ stems directly from the parameters
($N_{ij}$, $U_{ij}$, $V_{ij}$) of the chargino and neutralino
sectors. The cosmological bounds imposed by WMAP data on the $M_1$
parameter and its phase $\varphi_{U(1)}$ play an important role in
taking their shapes of the phase dependences of these processes.

In this study, we use the framework of R-parity conserving
supersymmetric scenarios wherein the lightest supersymmetric
particle (LSP) is a viable candidate for Cold Dark Matter (CDM).
Other than its relic density (observed by WMAP) little is known
about the structure of CDM. But the recent astrophysical
observations of the fluxes of high energy cosmic rays give
information about the properties of CDM. In particular, recent
results from Fermi LAT \cite{Abdo:2009zk} indicate an excess of the
electron plus positron flux at energies above 100 GeV. This also
confirms the earlier results from ATIC \cite{:2008zzr}. On the other
hand, PAMELA experiment \cite{Adriani:2008zr} reports a prominent
upturn in the positron fraction from 10-100 GeV, in contrast to what
is expected from high-energy cosmic rays interacting with the
interstellar medium. Although standard astrophysical sources such as
pulsars and microquasars may be able to account for these anomalies,
the positron excess at PAMELA and the electron plus positron flux of
Fermi LAT have caused a lot of excitement being interpreted as
decay/annihilation of Dark Matter. These unexpected results from
PAMELA, ATIC and Fermi LAT experiments imply a new constrain on LSP
that is CDM candidate: LSP must have not only the correct relic
density found by WMAP but also correct decay/annihilation rates into
electron-positron pairs. In the framework of the MSSM, a detailed
analysis of decay of CDM that includes the observed cosmic ray
anomalies has been given in Ref. \cite{Pospelov:2008rn}.

\end{document}